# Linearly polarized picosecond pulse shaping with variable profiles by a birefringent shaper

**Fangming Liu[1], Senlin Huang[1], Kexin Liu[1,*] and Shukui Zhang[2,*]**

[1]Institute of Heavy Ion Physics, School of Physics, Peking University, Beijing 100871, China
[2]Thomas Jefferson National Accelerator Facility, 12000 Jefferson Avenue, Newport News, VA 23606, USA

*Corresponding authors: *kxliu@pku.edu.cn* and *shukui@jlab.org*

**Abstract**

This paper describes the demonstration of linearly polarized picosecond pulse shaping with variable profiles including symmetric and non-symmetric intensity distributions. Important characteristics such as stability and transmission were studied, resulting in highly reliable performance of this fan-type birefringent shaping system. This variable temporal shaping technique is applicable over a wide range of laser parameters and may lead to new opportunities for many potential applications. A new double-pass variable temporal shaping method that significantly reduces the required crystal quantity is also proposed in this paper.

## 1. Introduction

Laser temporal shaping is an important technique, or a pivotal tool in some cases, to improve laser capability that otherwise may not be possible to certain applications. For example, parabolic pulses are highly desired in super-continuum generation (SCG) [1,2], ultra-short pulse generation [2,3], fiber amplifiers [4], optical regeneration [5], pulse retiming [6], spectral compression [7], and mitigation of linear waveform distortions [8], etc. Flattop pulses can significantly boost the efficiency of the optical parametric amplification (OPA) process [9]. Double pulses with tunable temporal spacing and amplitude ratio can efficiently improve laser ablation quality on metals [10]. Temporally shaped pulse trains can significantly improve the laser microprocessing quality for dielectrics [11,12]. In other areas such as time domain add-drop multiplexing [13], wavelength conversion [14], optical signal doubling [15], time-to-frequency mapping of multiplexed signals [16], etc., triangular or sawtooth pulses are often needed. The recent development in the electron accelerator-based light sources has pushed the performance of their photocathode drive lasers to an extreme that requires a flattop and even 3D ellipsoidal pulse intensity distribution in order to significantly reduce the emittance of electron beams [17,18].

There are different ways to generate a certain laser pulse profile, depending on the specific pulse length. Generally speaking, laser pulse shaping in the picosecond regime presents more challenges than pulses in other time regimes, due to the relatively short pulse duration and narrow spectral bandwidth of picosecond pulses. The acousto-optic or electro-optic shaping systems for nanosecond to millisecond pulse is too slow to work with picosecond laser pulses. In shorter time regime, the widely used spectral-domain shaping techniques such as grating-based shapers or acousto-optic programmable dispersive filter (DAZZLER) work very well with broadband femtosecond pulses but don't fit shaping picosecond pulses with narrow spectrum [19]. Although several methods have been demonstrated for narrow spectrum picosecond pulse shaping, involving the use of superstructured fiber Bragg gratings [20], interferometer-like structures with beam splitters [21] and cascaded birefringent crystals of different length with two orthogonally polarized combs of pulses [17], etc., none of these techniques allows for practical optical arbitrary waveform generation (OAWG) for narrowband picosecond laser pulses. Multiple birefringent crystals and polarizers with a Solc fan filter structure (referred to as fan-type hereafter) were successfully used for picosecond pulses shaping, but only flattop profile was demonstrated [22]. The fiber-based arbitrary temporal picosecond laser shaping technique reported recently also requires strongly-chirped broadband seed pulses and relies on fairly complicated time-frequency manipulation. Moreover, the liquid crystal controller tends to fail under high laser power [19]. Other methods for temporal shaping of picosecond lasers [9, 23] appear inflexible in practice as they require precise alignment of an array of different optical paths, bearing high degree of sensitivity to mechanical or thermal perturbations and are intended to produce only a few limited number of pulse shapes [24].

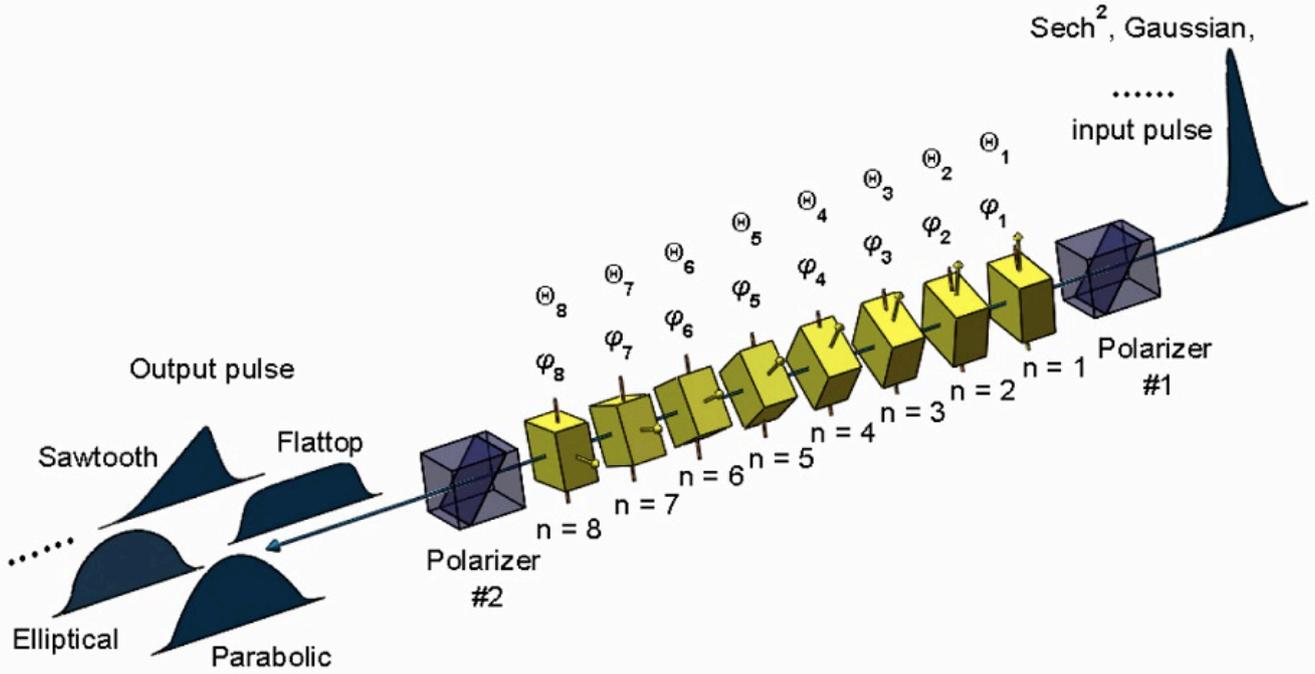

**Fig. 1.** The optical layout of the birefringent shaper used for variable pulse shaping. $\Theta_n$ is the angle between the slow axis of the *n-th* crystal and the polarization direction of the input polarizer #1. $\varphi_n$ is the crystal phase delay of the *n-th* crystal.

Stable picosecond pulse shaping with variable predefined shapes and linear polarization by another birefringent temporal shaper with a different type of configuration (referred to as folded-type hereafter) was recently demonstrated experimentally [25], where high transmittance and high stability were achieved simultaneously when crystal phase delay was set at $\pi$, while 0 rad crystal phase delay would significantly reduce the transmittance to nearly zero. In a further study as reported in this paper, we show variable picosecond pulse shaping with linear polarization can also be realized in a fan-type birefringent temporal shaper based on the Solc fan filter concept [26]. Such fan-type variable temporal shaping method shows distinctive characteristics compared with its peer (i.e., folded-type mentioned above) in that the high transmittance and high stability could be achieved simultaneously when crystal phase delay was set at 0 rad, with nearly zero transmittance when crystal phase delay was at $\pi$. As a matter of fact, birefringent system in a folded-type shaper operates in a half-wave-retarder (HWR) mode, while birefringent system in a fan-type shaper here operates in a full-wave-reatader (FWR) mode. The physical structure as well as the rules for tuning crystal rotation angles for laser temporal shaping between folded-type & fan-type birefringent temporal shapers are very different, but their performce in shaping the pulse profile are primarily the same. To our knowledge, this is the first time such fan-type temporal shaping method has been explored to such an extent that may lead to new opportunities for many potential applications. Similar to the folded-type variable shaping method shown in [25], the fan-type variable shaping method described below can also produce laser pulse with various predefined temporal profiles and linear polarization, and features important characteristics such as good long-term stability, easiness for operation and automation, and is directly applicable to shaping high power narrowband picosecond laser pulses over a wide laser wavelength range from IR to UV with any pulse repetition rate and any time structure.

## 2. Variable laser pulse shaper design

Figure 1 depicts the physical structure of the variable pulse shaper to be reported in this paper. Two polarizers labeled as #1 and #2 are alinged with their polarization directions parallel to each other. A number of identical birefringent stages are placed between polarizer #1 and #2. In such a shaper, any number of birefringent stages can be used though only eight birefringent stages (or eight birefringent crystals in our case) are depicted in Fig. 1. Each birefringent crystal is cut with its optic axis parallel to its end surfaces (also known as an a-cut crystal). The birefringent crystals are arranged in a way that the equation $\Theta_n = \frac{45°}{N}(2n-1)$ is satisfied, where $\Theta_n$ represents the angle between the slow axis of the *n-th* crystal and the polarization direction of the input polarizer #1, $N$ is the total number of the birefringent crystals and $n(=$

$1, 2, \cdots, N$) denotes the crystal sequence number. For a positive axial crystal (e.g., YVO4 crystal), its slow axis is the same as its optic axis, while for a negative axial crystal (e.g, BBO crystal), its slow axis is normal to its optic axis.

The simplest case in the birefringent crystal shaper is when only one birefringent crystal is used, which is also called single-stage fan-type shaper. Due to different group velocities of the ordinary (o) beam and extraordinary (e) beam in the birefringent crystal, the input laser pulse splits into two replica pulses corresponding to o beam and e beam respectively after travelling through the crystal. The time delay introduced between the two replica pulses is $T = l|1/v_{g,o} - 1/v_{g,e}|$, in which $l$ is the crystal length, $v_{g,o}$ and $v_{g,e}$ are the group velocities of o and e pulses in the birefringent crystal, respectively. The crystal phase delay, which determines the phase difference between the replica pulses, is defined as $\varphi = 2\pi(cT/\lambda - \lfloor cT/\lambda \rfloor)$, where $\lfloor \ \rfloor$ is a floor function used to round down to the next integer (e.g., $\lfloor 12.8 \rfloor = 12$), $c$ is the speed of light in vacuum, $\lambda$ is the laser wavelength. Upon exiting the output polarizer #2, the replica pulses interfere with each other, resulting a output pulse with intensity $I_{out}(t) \propto E_o^2(t) + E_e^2(t) + 2E_o(t)E_e(t)cos(\Delta\varphi')$ [27], where $E_o(t)$ and $E_e(t)$ refer to the electric fields of the two replica pulses corresponding to o beam and e beam, respectively. The $\Delta\varphi' = \varphi + 2\pi\lfloor cT/\lambda \rfloor$ is the phase difference between the two replica pulses. Due to interference, the output pulse intensity $I_{out}(t)$ strongly depends on the crystal phase delay $\varphi$ as an example shown in Fig. 2.

Due to the dependence of the birefringence upon temperature and the thermal expansion of the crystal, the crystal phase delay can be fine-tuned between $0 \sim 2\pi$ by adjusting the temperature of the crystal. This makes the birefringent crystal a temperature-controlled variable wave retarder.

In general, a shaper contains $N$ birefringent crystals would produce $2^N$ mutually delayed replica pulses. Under the condition that all the crystals are of equal thickness and introduce the same crystal time delay $T$, the $2^N$ replica pulses exiting the shaper can be arranged into $N + 1$ groups to form $N + 1$ replica pulses equally spaced in time [28]. Increasing $\Theta_n$ would increase the amplitude of the $(N + 2 - n)$-th replica pulse and decrease the amplitude of the $(N + 1 - n)$-th replica pulse, meanwhile giving very small influence on the amplitudes of all other $N - 1$ replica pulses, and vice versa. Based on this rule, by varying the rotation angles $\Theta_n$ of the crystals, the relative amplitudes of these $N + 1$ replica pulses can be flexibly adjusted to generate output pulses with basically any predefined shape.

In this work, the input and the output polarizers are two Glan polarizers with an extinction ratio on the order of $10^5$:1. Eight 3.3mm a-cut Yttrium Orthovanadate (YVO4) crystals were employed for shaping 532nm picosecond laser pulses. The end surfaces of each crystal were anti-reflection coated at 532nm with residual reflection about 0.1%. Each crystal can be independently rotated around the laser beam propagation axis through a kinematic rotation stage which also provides adjustments of pitch and yaw. The crystals were housed in crystal ovens, which allow the temperature of each crystal to be independently controlled between 40°C ~ 180°C with an accuracy of 0.1°C.

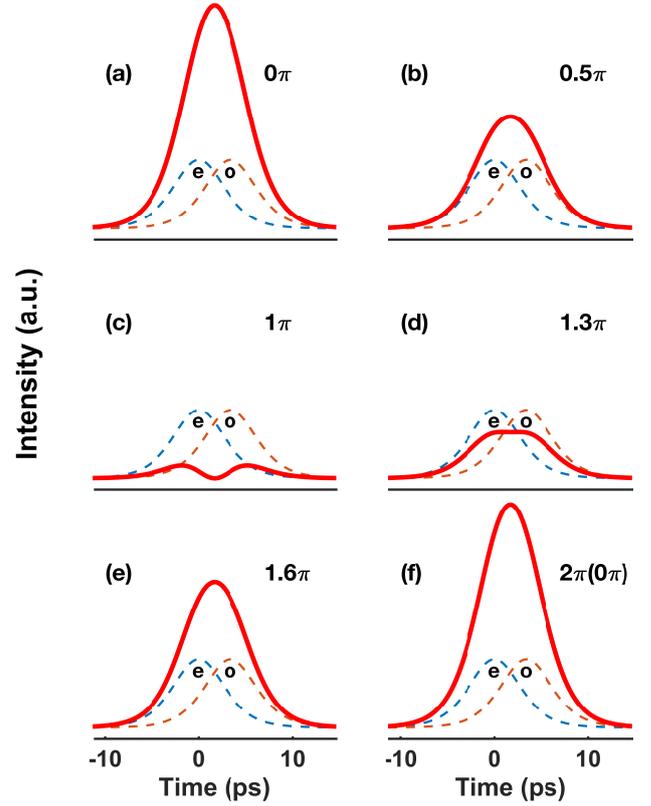

Fig. 2. Calculated output pulse intensity profiles (red solid lines) of a single-stage fan-type shaper for crystal phase delay at (a) $0\pi$, (b) $0.5\pi$, (c) $1\pi$, (d) $1.3\pi$, (e) $1.6\pi$, and (f) $2\pi$ or $0\pi$, respectively. The blue and orange dashed lines represent the intensity profiles of replicas corresponding to e beam and o beam, respectively. All the curves are normalized with respect to the output pulse intensity at 0 rad crystal phase delay. The input pulse is a 532nm laser pulse with Sech$^2$ temporal profile and 6.5ps (FWHM) pulse width. The birefringent crystal is a 3.3mm long a-cut YVO4 crystal.

## 3. Laser shaping experiment and results

The experiment setup for generation and measurement of the shaped pulse has been described in detail in [25], where the envelope of the shaped pulses were measured in real time with sub-picosecond temporal resolution by a cross-correlator. The 532nm laser pulse used for shaping has a near Sech$^2$ temporal profile with 6.5ps (FWHM) pulse width and 8.125MHz pulse repetition rate.

The temperature calibration for crystal phase delay setting can be carried out using the afore-mentioned single-stage fan-type shaper. When the crystal rotation angle is set to 45°, the output pulse envelop of the single-stage shaper would be a symmetrical profile. When tuning the crystal phase delay by adjusting the crystal temperature, both the intensity and shape of the output pulse would change in a way as shown in Fig. 2, which can be monitored by the cross-correlator and laser power meter. When the crystal phase delay is tuned to π, the two replica pulses interfere destructively with each other, resulting in a double-peak output pulse of smallest intensity with its center drooping down to zero. At this point, the birefringent crystal becomes a HWR as employed by folded-type shaper in [25]. As the crystal phase delay shifts away from π, the intensity of the output pulse grows and reaches the maximum through constructive interference when crystal phase delay gets 0 rad, i.e., the birefringent crystal now becomes a FWR. In the experiment, a 3.3mm YVO$_4$ crystal requires a temperature variation $\Delta T_o$ of about 36°C in order to shift the crystal phase delay by 2π for 532nm laser.

To achieve both high transmittance and high stability, the temperature of each crystal was carefully adjusted to set all the crystal phase delays close to 0 rad through the above-mentioned method before shaping the laser pulse, as the initial crystal phase delays of eight YVO$_4$ crystals are randomly distributed between 0 ~ 2π. This makes each crystal a FWR. The output pulse of this fan-type shaper varies periodically with the crystal phase delay at an interval of 2π. Within one cycle, the fan-type shaper's transmittance declines when increasing or decreasing all the crystal phase delays away from 0 rad, and would even drops to nearly zero level if all the crystal phase delays are close to π.

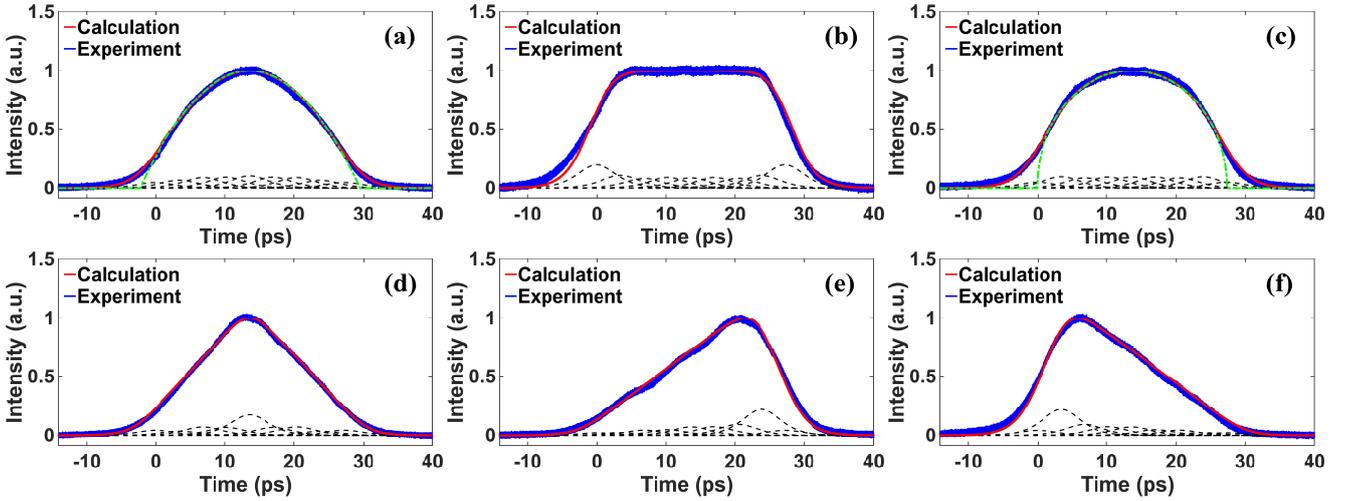

**Fig. 3.** Measured (blue lines) and the calculated (red solid lines, corresponding crystal rotation angle corrections are listed in Table 1) pulse profiles after the shaper, (a) Parabolic, (b) Flattop, (c) Elliptical, (d) Triangular, (e) Sawtooth-I, (f) Sawtooth-II. Black dashed curves are theoretically calculated 8+1 replica pulses. Green dash-dotted lines are the ideal parabolic and elliptical curves. Birefringent stages in the shaper are identical. See detailed description in the text.

**Table 1.** Calculated crystal rotation angle corrections $\Delta\Theta_n$ for different output pulse shapes, where $\Delta\Theta_n = \Theta_{n,tuned} - [\frac{45°}{N}(2n-1)]$, $\Theta_{n,tuned}$ represents the crystal rotation angles of the calculated output pulse shapes. $\Delta\Theta_n$ is obtained by trial-and-error method according to the rule described in the section prior to Fig. 2. All the crystals in the shaper are a–cut YVO$_4$ with identical thickness of 3.3mm and phase delay set to 0 rad. The input pulse profile is a Sech$^2$ distribution, with 6.5ps FWHM pulse width and 532nm wavelength.

| Different shapes \ Crystal sequence number $\Delta\Theta_n(Deg)$ | 1 | 2 | 3 | 4 | 5 | 6 | 7 | 8 |
|---|---|---|---|---|---|---|---|---|
| Parabolic | 0.2 | -2.6 | -2.65 | -1.2 | 1.2 | 2.65 | 2.6 | -0.2 |
| Flattop | 4 | 2 | 0.5 | 0.1 | -0.1 | -0.5 | -2 | -4 |
| Elliptical | 0.1 | -1.7 | -2.05 | -0.85 | 0.85 | 2.05 | 1.7 | -0.1 |
| Triangular | 0.5 | -3.5 | -3.8 | -2.8 | 2.8 | 3.8 | 3.5 | -0.5 |
| Sawtooth-I | 0.5 | 4.5 | 6.5 | 8.4 | 9.9 | 9.1 | 6.6 | 1.2 |
| Sawtooth-II | -1.2 | -6.6 | -9.1 | -9.9 | -8.4 | -6.5 | -4.5 | -0.5 |

By following the procedures described in [25], several predefined target pulse shapes were generated simply through tuning the crystal rotation angles with the help of a real time cross-correlator, as shown in Fig. 3 where smooth profiles of shaped pulses can be seen. The measured results agree well with the calculated profiles, indicating the effectiveness of this shaping method for realizing variable temporal profile shaping. Table 1 gives the crystal rotation angle corrections for the calculated pulse profiles. The smoothness of the pulse shapes can be primarily attributed to the relatively large ratio of input pulse width $T_1$ (6.5ps) versus crystal time delay $T$ (3.4ps), as well as the 0 rad crystal phase delays for all crystals. The total transmittance of the shaper depends on the output pulse shapes, and typically is about 20%. Taking into account the absorption and reflection loss of each element in the system, the total transmittance could be up to 30%, and our calculation shows it is possible to reach 40~50% for most pulse shapes under optimal condition. We believe factors such as the imperfections of elements, etc., may be responsible for such difference. The intensity modulation in the flattop region of the measured flattop pulse is about 0.35% (rms), which may be further reduced through a finer optimization of the crystal rotation angles. The rising and falling edges of the shaped pulses are primarily determined by the pulse width and profile of the input pulse. One way to make the profile edge of the shaped pulse closer to an ideal geometric profile is to use shorter input pulse and more crystals although the pulse shapes in Fig. 3 are satisfactory for many practical applications.

The stability of the shaped pulse, which is important for practical applications, is also investigated in the experiment. During an 8-hour continuous operation, the pulse shape stayed nearly constant as an example can be seen from the recorded data in Fig. 4(a) for a triangular pulse. The beam transmittance through the shaper also remained unchanged, staying at about afore-mentioned 20% level. This indicates that the shaper has an excellent long-term stability. Further analysis reveals that this high stability can be mainly attributed to the fact that the crystal phase delays were fine controlled and were all set to 0 rad by precisely controlling the crystal temperature. Figure 4(b) shows the calculated triangular pulse profiles with crystal phase delay equally spaced between $0\pi$ and $0.1\pi$. The interval of crystal phase delay between two adjacent profiles is $0.02\pi$, which correspondes to a 0.36°C temperature variation. As the crystal phase delay shifts away from $0\pi$ with the same amount of temperature variation, the relative peak intensity change between two adjacent profiles are about 0.7%, 2.1%, 3.4%, 4.7%, and 6%, respectively. This in turn shows that the influence of the temperature variation to the shaped pulse declines rapidly and become negligible as the crystal phase delay approaches 0 rad. The high stability obtained at 0 rad crystal phase delay stems from the nature of coherent light addition. For the output pulse of the shaper, it can be regarded as the coherent addition of $2^N$ replica pulses of the input laser pulse. When any two of the replica pulses with intensity of $I_j$ and $I_k$ combined coherently, the resulting light intensity is $I_{jk} = I_j + I_k + 2\sqrt{I_j I_k}\cos(\varphi'_j - \varphi'_k)$, where $\varphi'_j$ and $\varphi'_k$ refer to the phase of replica pulse $j$ and replica pulse $k$, respectively. Due to interference, the intensity $I_{jk}$ strongly depends on the phase difference $\Delta\varphi' = \varphi'_j - \varphi'_k$, which is closely related to the crystal phase delay. When setting all the crystal phase delays to 0 rad (note, $\pi$ is not selected since it leads to nearly zero beam transmittance through the shaper), the phase difference $\Delta\varphi'$ would become an integer number of $\pi$ for all the replica pulses, and results in very small, nearly zero functional slope $\frac{\partial I_{jk}}{\partial \Delta\varphi'}$. Therefore, at this point the shaper becomes most tolerant to the impact of temperature fluctuation which causes $\Delta\varphi'$ to change.

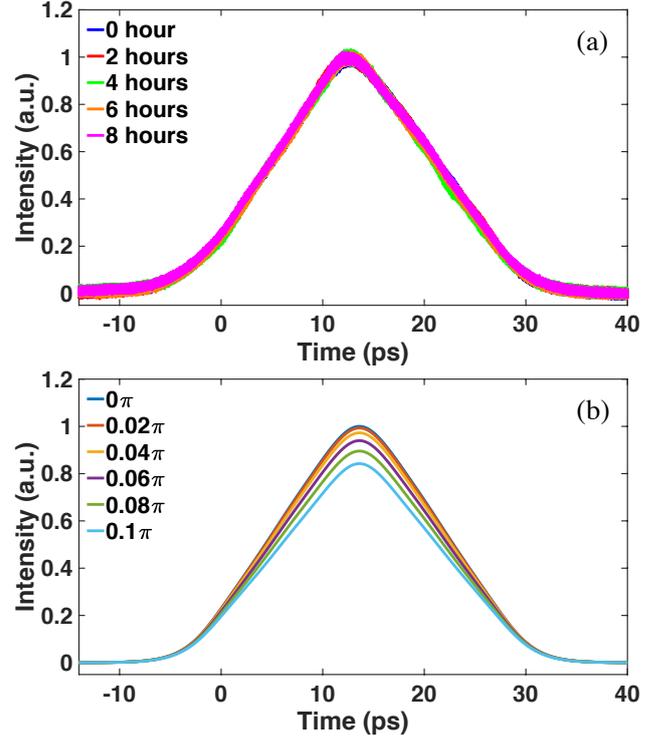

**Fig. 4.** (a) Recording of measured triangular pulse shapes at different times during 8-hour continuous operation when each crystal phase delay was set to 0 rad. (b) Calculated triangular pulse profiles with all the crystal phase delays set to $0\pi$, $0.02\pi$, $0.04\pi$, $0.06\pi$, $0.08\pi$, and $0.1\pi$, respectively. The specific crystal rotation angle corrections for the calculated triangular pulse profiles are listed in Table 1. The crystals and the input pulse are the same as those in Table 1.

## 4. Further discussion

Besides the birefringent crystals, the birefringent stages in the above-mentioned fan-type and folded-type shapers may also be realized by electro-optic retarders [29] or fiber components [30,31], etc., which would further extend the application range of such variable temporal shaping technique. According to the calculation, higher transmittance of the shaper can also be reached by increasing the ratio of input pulse width over crystal time delay $T_1/T$. On the other hand, decreasing the ratio $T_1/T$ would enable such shaper to produce tunable pulse trains or pulses with ripples.

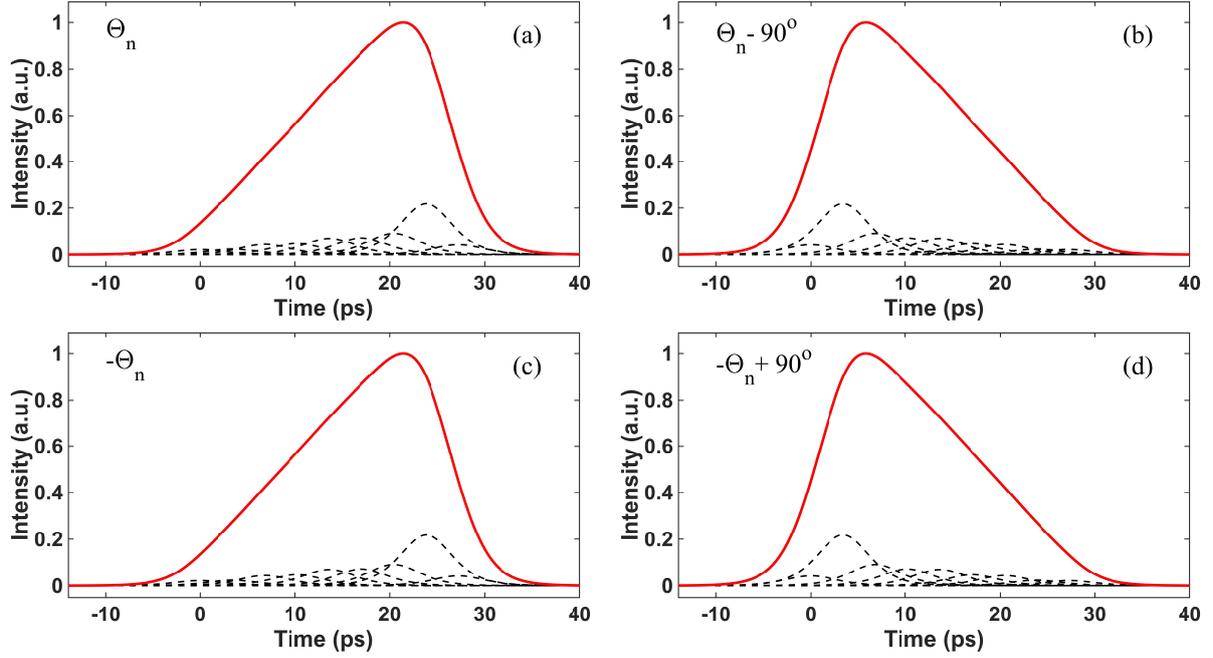

**Fig. 5.** Calculated (red solid lines) sawtooth pulses of the shaper with crystal rotation angles set at $(a)$: $\Theta_n$, $(b)$: $\Theta_n - 90^o$, $(c)$: $-\Theta_n$, and $(d)$: $-\Theta_n + 90^o$, respectively. The specific crystal rotation angle values for different sawtooth pulses in this figure are listed in Table 2. Black dashed curves are the theoretically calculated 8+1 replica pulses. The crystals and the input pulse are the same as those in Table 1.

**Table 2.** Calculated crystal rotation angles ($\Theta_{n,tuned}$) for sawtooth pulses from the shaper with crystal rotation angles set at $\Theta_n$, $\Theta_n - 90^o$, $-\Theta_n$, and $-\Theta_n + 90^o$, respectively. The crystals and the input pulse are the same as those in Table 1.

| Crystal rotation angles $\Theta_{n,tuned}$(Deg) / Crystal sequence number | $\Theta_n$ | $\Theta_n - 90^o$ | $-\Theta_n$ | $-\Theta_n + 90^o$ |
|---|---|---|---|---|
| 1 | 6.125 | 6.125 − 90 | −6.125 | −6.125 + 90 |
| 2 | 21.375 | 21.375 − 90 | −21.375 | −21.375 + 90 |
| 3 | 34.625 | 34.625 − 90 | −34.625 | −34.625 + 90 |
| 4 | 47.775 | 47.775 − 90 | −47.775 | −47.775 + 90 |
| 5 | 60.525 | 60.525 − 90 | −60.525 | −60.525 + 90 |
| 6 | 70.975 | 70.975 − 90 | −70.975 | −70.975 + 90 |
| 7 | 79.725 | 79.725 − 90 | −79.725 | −79.725 + 90 |
| 8 | 85.575 | 85.575 − 90 | −85.575 | −85.575 + 90 |

Based on our simulations, the fan-type shaper introduced above may operate in several different configurations which are related to each other. For example, if a shaper has $-\Theta_n$ crystal rotation angle setting, its output pulse would be the same with that from the shaper with $\Theta_n$ angle setting, as shown in Figs. 5(a) and 5(c), or Figs. 5(b) and 5(d), respectively. While for a shaper with $\Theta_n \pm 90^o$ angle setting, its output pulse would be a mirror image of that from the shaper with $\pm\Theta_n$ angle setting, as shown in Figs. 5(a) and 5(b), or Figs. 5(c) and 5(d), respectively.

From Table 1, we can see that the single-pass shaper shown previously in Fig. 1 can be transformed into a new type of double-pass shaper as shown in Fig. 6, which produces the same symmetrical pulses as those from the single-pass shaper. In a double-pass shaper, only the first half section of the single-pass shaper is needed, while the second half section can be simply replaced by a retro-reflector and an quarter-wave-retarder (QWR). The QWR is aligned with its fast or slow axis oriented at 45° with respect to the polarization direction of the input polarizer. The combination of retro-reflector and QWR acts as a half-wave-retarder (HWR) inducing the polarization rotation for all the replica pulses and returning the laser beam back on the same beam path. On the return pass, the replica pulses see the shaper crystals the same way as they would in the second half group of the crystals in a single-pass configuration. The shaped pulse would then be conveniently kicked out by the input polarizer upon its return, with its polarization direction perpendicular to that of the input pulse. Since the laser beam passes through each crystal twice, the required crystal quantity can be reduced by half, rendering much more simplicity and reducing the cost.

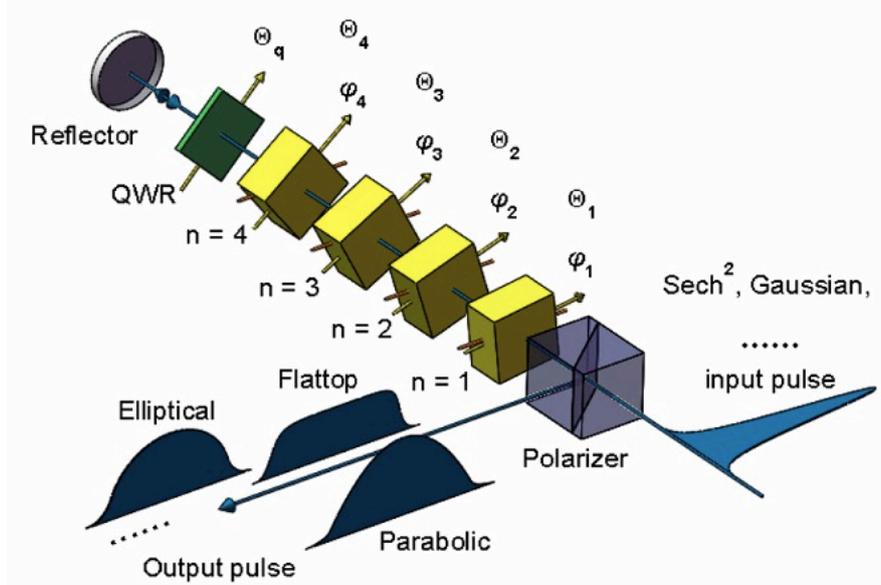

**Fig. 6.** The layout of a double-pass birefringent shaper. $\Theta_n$ is the angle between the slow axis of the *n-th* crystal and the polarization direction of the input polarizer. $\varphi_n$ is the crystal phase delay of the *n-th* crystal. $\Theta_q$, representing the angle between the fast or slow axis of the QWR and the polarization direction of the input polarizer, is 45° for this double-pass shaper.

The fan-type variable temporal shaping method in this paper has many distinctive features potentially for broad applications, just like the folded-type variable temporal shaping method in [25]. In addition to the differences described above, actually all the other features of the folded-type shaper are also applicable to this fan-type shaper, making these two kinds of shapers interchangeable in many applications.

In principle, both the above-mentioned folded-type and fan-type shaping methods can be scaled approximately against the ratio $T/T_1$ to shape laser pulses with initial pulse duration in femtosecond and nanosecond regimes. Both of these two types of shaping model are based on the assumption that the replica pulses do not change significantly when travelling through the elements of the shaper, and are suitable for shaping picosecond pulse. For very long input pulse such as nanosecond pulse, though with very narrow spectrum bandwidth, other kinds of birefringent stages are needed to replace the bulk birefringent crystals. This is bacause of that shaping for nanosecond pulse with birefringent crystal would be limited by the requirements of very long crystal of up to meters to provide the required time delay $T$, as well as ultrahigh-precision temperature control system since the $\Delta T_0$ (amount of temperature variation for shifting the crystal phase delay by a cycle of $2\pi$) is approximately inversely proportional to the crystal length [25]. The folded-type and fan-type shaping methods may be directly applied to shaping sub-picosecond pulses where the effect of dispersion is not much a concern. However, for ultrashort femtosecond laser pulse with broadband, the dispersion and nonlinear intensity-dependent effects in the optical materials would significantly stretch or distort the replica pulses. This is an interesting topic for future research currently under our consideration.

## 5. Summary

In this paper, generation of linearly polarized picosecond pulses with variable predefined profiles is demonstrated both theoretically and experimentally through coherent addition shaping in a fan-type birefringent shaper. Several interesting pulse shapes including parabolic, flattop, elliptical, triangular, and sawtooth shapes were experimentally produced. It is discovered that high transmittance and high stability can be achieved simultaneously by setting all crystal phase delays to

0 rad through fine control of the crystal temperature. The measured transmittance of the shaper is ~20%, which can be further improved according to the calculation. Similar to the folded-type shaper [25], this fan-type shaper is also robust, easy to use and be automated, capable of shaping laser pulses over a wide laser wavelength range from IR to UV with any pulse repetition rate, any time structure and high power. Linear polarization makes the shaped pulse possible to be applied to a wide range of applications, such as laser amplifiers, nonlinear frequency conversion, interferometers, polarization-dependent devices (e.g., Faraday rotators, Pockels cells), high brightness photocathode electron sources, and so on. Finally, a promising double-pass variable temporal shaping method is proposed and could significantly reduce the required crystal quantity with much more simplicity and lower cost.

**Funding**

National Key Research and Development Program of China (Grant No. 2016YFA0401904 and Grant No. 2017YFA0701000); National Natural Science Foundation of China (Grant No. 11735002).